\documentclass[pra,aps,twocolumn,10pt,floatfix,superscriptaddress,showpacs]{revtex4-1}
\usepackage{amsmath,bm,graphicx}
\usepackage{amssymb}
\usepackage{amsfonts}
\usepackage{color}
\usepackage{subfigure}

\begin{document}
\def \figwidth {\columnwidth}

\title{A nonlinear least squares method for the inverse droplet coagulation problem}
\author{Peter P. Jones}
\affiliation{Centre for Complexity Science, University of Warwick, Coventry CV4 7AL, UK}

\author{Robin C. Ball}
\affiliation{Dept.~of Physics, University of Warwick, Coventry CV4 7AL, UK}
\affiliation{Centre for Complexity Science, University of Warwick, Coventry CV4 7AL, UK}

\author{Colm Connaughton}
\email{connaughtonc@gmail.com}
\affiliation{Mathematics Institute, University of Warwick, Coventry CV4 7AL, UK}
\affiliation{Centre for Complexity Science, University of Warwick, Coventry CV4 7AL, UK}

\date{\today} 
 
\begin{abstract}
If the rates, $K(x,y)$, at which particles of size $x$ coalesce with particles of size $y$ is known, then the mean-field evolution of the particle-size distribution of an ensemble of irreversibly coalescing particles is described by the Smoluchowski equation. We study the corresponding inverse problem which aims to determine the coalescence rates, $K(x,y)$ from measurements of the particle size distribution.  We assume that $K(x,y)$ is a homogeneous function of its arguments, a case which occurs commonly in practice. The problem of determining, $K(x,y)$, a function to two variables, then reduces to a simpler problem of determining a function of a single variable plus two exponents, $\mu$ and $\nu$, which characterise the scaling properties of $K(x,y)$. The price of this simplification is that the resulting least squares problem is nonlinear in the exponents $\mu$ and $\nu$. We demonstrate the effectiveness of the method on a selection of coalescence problems  arising in polymer physics, cloud science and astrophysics. The applications include examples in which the particle size distribution is stationary owing to the presence of sources and sinks of particles and examples in which the particle size distribution is undergoing self-similar relaxation in time.
\end{abstract}

\maketitle

\section{Introduction}

\paragraph*{}
Coagulation processes abound in nature and span all scales, ranging from the microscopic scales of atmospheric aerosol formation \cite{Friedlander1977}, to the cosmological scales of the clustering of matter within the universe \cite{SilkWhite1978}. A particular example which played a considerable role in motivating this work is droplet coalescence in clouds. The role of droplet coalescence in the formation and internal dynamics of clouds is of considerable contemporary interest. This is because improved understanding of the evolution of the droplet size distribution in clouds would increase the precision of climate evolution projections \cite{STE2005}. Much current research in this area focuses on determining the rate of coalescence between droplets of different sizes. Turbulence in the cloud air mass complicates this task significantly. It plays a non-trivial role in determining the collision rate of water droplets \cite{BMSS2010,GrabWang2009,grabowski_growth_2013}. Direct numerical simulation of the 
dynamics of droplets in 
turbulent flows \cite{RC2000,WARG2008} are possible. It is not clear however that these simulations can yet span the range of scales required to obtain a full understanding of the role of turbulence in facilitating droplet collisions in a cloud \cite{onishi_warm-bincold-bulk_2012}. For these reasons an a-priori understanding of droplet collision rates in clouds remains elusive. Recent technological advances, however, have improved both the quality and quantity of empirical data on droplet size distributions at various stages of cloud evolution \cite{SLWF2006}. It is therefore timely to address the possibility of using these observations to solve the inverse problem of determining collision rates from measurements of the droplet size distribution. This is the topic of this article. We aims to develop a data-driven approach to determining collision rates which can complement the insights emerging from current theoretical and numerical work on this problem. While we have motivated our work with in the context 
of droplet coagulation in clouds the 
methods which we develop are quite general and provide a  quantitative means of constraining the choice of model in any coagulation problem in which the microphysics is unknown or controversial.

\paragraph*{}
Throughout this article, we characterise the size of droplets by their mass, $m$, and use $N(m,t)$ to denote the droplet size distribution at time $t$. We denote the rate of coalescence between droplets of sizes $m_1$ and $m_2$ by $K(m_1,m_2)$. This function is often referred to as the ``collision kernel''.
We assume throughout that the solution of the forward problem of determining $N(m,t)$ if $K(m_1,m_2)$ is known is obtained by solving the Smoluchowski coagulation equation (SCE) \cite{Smol1917a}:
\begin{align}
\nonumber \partial_t N_m(t) &= \frac{1}{2}\int_0^{m}dm_1\, K(m_1,m-m_1) N_{m_1}(t) N_{m-m_1}(t)\\
\label{eq-Smoluchowski}  &- N_m(t) \int_0^{M}dm_1\, K(m,m_1)\, N_{m_1}(t)\\
\nonumber &+ \frac{J}{m_0}\,\delta(m-m_0),
\end{align}
together subject to an initial condition $N_m(0)=\mathcal{N}_0(m)$. The final term described a source of particles which injects ``monomers'' having mass $m_0$ at a rate $J$. Depending on the application, $J$ could be zero. We shall take the smallest droplet size in the system to be $m_0=1$. We have also explicitly introduced a cutoff mass, $M$. Droplets larger than $M$ are removed from the system. Depending on the application, $M$ could be infinite.
 The forward problem is nonlinear in the unknown, $N(m,t)$. The inverse problem which forms the topic of this article is to determine $K(m_1,m_2)$ from measurements $N(m,t)$. Note that the inverse problem is linear in the unknown, $K(m_1,m_2)$. The non-triviality of the inverse problem comes from the fact that, since we seek to determine a function of two variables from a function of one variable, we would generally expect it to be ill-posed. 

Notable previous work on this problem includes  the work of Onishi and coworkers in the atmospheric science context \cite{OMTKK2011} and the work of Ramkrishna and coworkers in the chemical engineering context \cite{muralidar_inverse_1986,muralidhar_inverse_1989,WR1992}. Onishi et al \cite{OMTKK2011} address the problem of ill-posedness by using significant prior knowledge about droplet coalescence in turbulent conditions to put strong constraints on the functional form of the kernel. Specifically it was assumed that the collision rate could be modeled as a linear superposition of the gravitational sedimentations and Saffmann-Turner kernels. This simplified the inverse problem to a parameter estimation problem at the expense of a loss of generality. On the other hand, the methods pioneered by Ramkrishna et al \cite{WR1992} do not strongly constrain the functional form of $K(m_1,m_2)$. These authors address the problem of 
ill-posedness using a procedure known as Tikhonov regularisation.  In a previous note \cite{connaughton_remarks_2011} we explored the ability of the method in \cite{WR1992} to solve the inverse problem for kernels of the form, $K(m_1,m_2) = \frac{1}{2}(m_1^{\lambda} + m_2^{\lambda})$ with $0 \leq \lambda \leq 1 $. It was found that the method performed relatively poorly when the exponent $\lambda$ was fractional. This stems from the fact that $K(m_1,m_2)$ was represented using Laguerre polynomials which contain only integer powers.

\paragraph*{}
In this paper, motivated by the fact that many practical collision kernels contain fractional powers, we present a refined 
inverse method which deals with the problem of fractional exponents up front. Our method splits the problem into two stages. In the first stage we solve a (nonlinear) parameter estimation problem to determine a pair of exponents which best describes how $K(m_1,m_2)$ behaves for large and small masses. In the second stage we solve a (linear) inverse problem which uses the observations of $N(m,t)$ to correct the detailed form of $K(m_1,m_2)$ without changing the scaling exponents determined in the first stage. In order to simplify our task, we restrict ourselves to cases in which the collision kernel is a homogeneous function of its arguments. While this is common in practice, it is a weakness of our approach as compared with that of 
\cite{OMTKK2011} which does not require this restriction. 

We test our method on two broad classes of inverse problems. The first class consists of stationary
problems.  These occur when a source particles is present, $J>0$ in Eq.\eqref{eq-Smoluchowski}, and the sink at $M \gg m_0$ is important. For sufficiently large times, the droplet size distribution becomes independent of time \cite{Hayakawa1987} and describes a flux of mass through the space of droplet sizes from the injection scale, $m_0$, to the sink scale, $M$ \cite{CRZ2004stat}. The second class consists of time-evolving problems without a source of particles in which the droplet size distribution relaxes from a prescribed initial
condition which we take to be monodisperse. For such problems, homogeneous collision kernels usually result in the droplet size distribution
becoming self-similar for large times provided the characteristic size remains smaller than $M$.

\paragraph*{}
The remainder of the article is laid out as follows. In Sec.\ref{sec:kernels} we introduce some properties kinds of the kind of collision kernels we expect to find in applications and discuss ways of representing such kernels mathematically. Next we discuss various aspects of the forward problem relevant to our subsequent discussion of the inverse problem. The time-dependent forward problem is outlined in Sec.\ref{sec:timeDependentForwardProblem} and the stationary forward problem in Sec.\ref{sec:statdists}. Our main results on the inverse problems are presented in Sec.\ref{sec:retrievestatdistkern} and Sec.\ref{sec:scalingDecayRetrieval} for the stationary and time-dependent inverse problems respectively. Finally, \S\ref{sec:conclusionsAndOutlook} presents our conclusions and suggestions for further 
work.

\section{Representations of homogeneous collision kernels}
\label{sec:kernels}

\paragraph*{}
In this paper we focus on scale invariant problems for which the kernel, $K(m_1,m_2)$ is a homogeneous symmetric function of its arguments. We denote the overall degree of homogeneity by $\lambda$:
\begin{align}
 K(am_1,am_2) & = a^{\lambda}K(m_1, m_2).
\end{align}
Such kernels are important because many physical aggregation processes exhibit homogeneity for some range of scales \cite{KangRedner1986}. The following model kernel, primarily used in the analysis of scaling solutions of the SCE \cite{DongenErnst1988,Leyv2003}, will be of particular importance to us:
\begin{align}
 \label{eq-symmkernel} K_0(m_1, m_2) & = \frac{g}{2}\left(m_1^{\mu}m_2^{\nu} + m_2^{\mu}m_1^{\nu}  \right).
\end{align}
where $g$ sets the overall amplitude. Clearly we must have $\lambda = \mu + \nu$. For convenience, we adopt the convention throughout that $\nu \geq \mu $. The exponents  $\mu$ and $\nu$ then
capture the behaviour of the kernel Eq.~(\ref{eq-symmkernel}) when one mass is much larger than the other:
\begin{align}
   K_0(m_1, m_2) & \sim m_1^{\mu}m_2^{\nu}, \qquad \text{with~} m_1 \ll m_2. 
\end{align}
Most kernels occuring in practice are not of the form (\ref{eq-symmkernel}). One can however usually \footnote{One could imagine a kernel involving, for example, logarithmic mass dependences for which this is not true but we are not aware of such functional forms arising in practice.} assign a value to the exponents $\mu$ and $\nu$ by analogy with Eq.~(\ref{eq-symmkernel}) by considering the behaviour in the limit $m_1 \ll m_2$. We can then write any kernel as a product of (\ref{eq-symmkernel}) and another function, $F(m_1, m_2)$:
\begin{align}\label{eq-kerneldecomposition}
    K(m_1,m_2)  & = K_0(m_1, m_2)\,F(m_1, m_2).
\end{align}
Since $\lambda=\mu+\nu$, $F$ must be homogeneous of degree zero. It can therefore be expressed as a function, $f$, of a single variable $x = m_1/m_2$. 
\begin{align}
\label{eq-SF}
 F(m_1,m_2) &= f \left( \frac{m_1}{m_2} \right).  
\end{align}
Since $K(m_1,m_2)$ and, by extension, $F(m_1,m_2)$ is a symmetric function of its arguments, $f$ must have the symmetry:
\begin{align}
  f(x) = f \left(x^{-1} \right). 
\end{align}

\begin{figure}[tb]
\includegraphics[width=0.5\textwidth,keepaspectratio=true]{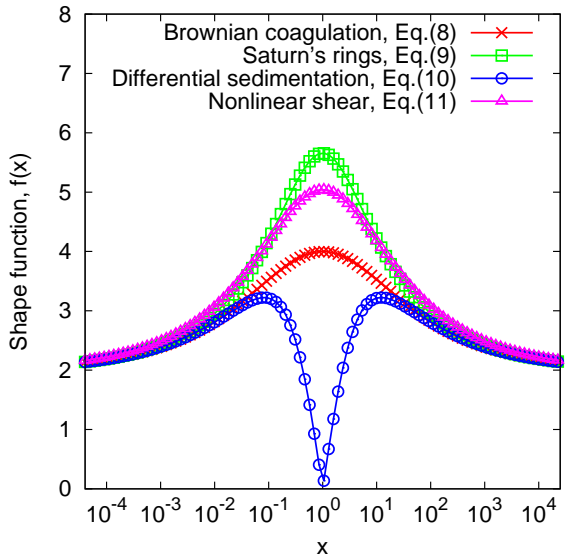}
\caption{The shape functions, as defined by Eqs.\eqref{eq-kerneldecomposition} and \eqref{eq-SF}, for the Brownian coagulation kernel, Eq.\eqref{eq-KBC}, the astrophysical coalescence kernel, Eq.\eqref{eq-KSR}, the differential sedimentation kernel, Eq.\eqref{eq-KDS}, and the nonlinear shear velocity kernel, Eq.\eqref{eq-KNLSV}.}
\label{fig-shapefunctions} 
\end{figure}

\paragraph*{}
We will refer to $f(x)$ the \textit{shape function} of the kernel. Since it is a homogeneous function of degree zero it is "almost" a constant and by construction must asymptote to a constant value as $x \to 0$ and $x\to \infty$. For the sake of concreteness, we have selcted the following kernels from the literature to use as test problems in this paper:  
\begin{eqnarray}
\label{eq-KBC} K_{\text{BC}}(m_1,m_2) & =& (m_1^{\frac{1}{3}} + m_2^{\frac{1}{3}})(m_1^{-\frac{1}{3}} + m_2^{-\frac{1}{3}}) \\
\label{eq-KSR} K_{\text{SR}}(m_1,m_2) &=&  (m_1^{\frac{1}{3}} + m_2^{\frac{1}{3}})^2(m_1^{-1} + m_2^{-1})^{\frac{1}{2}}\\
\label{eq-KDS} K_{\text{DS}}(m_1, m_2) & =& \left(m_1^{\frac{1}{3}} + m_2^{\frac{1}{3}} \right)^2 \left| m_1^{\frac{2}{3}} - m_2^{\frac{2}{3}} \right|\\
\label{eq-KNLSV} K_{\text{SNLV}}(m_1, m_2) & =& \left( m_1^{1/3} + m_2^{1/3}\right)^{7/3}.
\end{eqnarray}
Eq.\eqref{eq-KBC} is the kernel for Brownian coagulation of spherical droplets \cite{Smol1917a}. It has $\nu=1/3$ and $\mu=-1/3$. Eq.\eqref{eq-KSR} is the kernel describing aggregation of ice clusters due to differential orbital speed in planetary rings \cite{BBK2009,SmitEtAl1994}. It has $\nu=2/3$ and $\mu=-1/2$. Eq.\eqref{eq-KDS} is the kernel most relevant for the cloud problems since it describes coalescence of spherical droplets undergoing differential sedimentation in the Stokes regime in still air \cite{pruppacher_microphysics_1997}.  It has $\nu=4/3$ and $\mu=0$. Eq.\eqref{eq-KNLSV} is the so-called nonlinear velocity kernel describing shear-driven coagulation \cite{Aldous1999,SmitEtAl1994}. It has $\nu=7/9$ and $\mu=0$. The respective shape functions for each of these kernels are shown in \figurename{\ref{fig-shapefunctions}}. Note, how these shape functions are all asymptotically constant, and
despite the seeming functional complexity of the original kernels, have a rather simple form.

Our approach to solving the inverse problem described in the introduction will be to first estimate the exponents $\mu$ and $\nu$ and then correct the result by an appropriate shape function, $f(x)$. Since the shape function is a function of one variable, this should be a considerably easier problem. This is very much in the spirit of the original work of Ramkrishna and coworkers \cite{muralidar_inverse_1986, muralidhar_inverse_1989} who also exploited the fact that a homogeneous function of two variables is determined by its degree and an auxialliary function of a single variable. Our approach is an improvement in the sense that the function, $f(x)$, which we
need to determine is asymptotically constant at large and small values making it much 
easier to deal with.

To proceed we will need a way of representing functions on the interval $\left[M^{-1}, M\right]$ (remember we have taken $m_0=1$) which have the symmetry $f(x)=f(x^{-1})$. As is evident from Fig.\ref{fig-shapefunctions}, any symmetric function of $\log x$ has the required property. That is we take
\begin{align}
  f(x) & = h\left( \frac{\pi\,\log(x)}{\log(M)}\right) \label{eq-hy}
\end{align}
where $h(y)$ is any symmetric function on the interval $\left[-\pi,\pi\right]$. We
enforce symmetry indirectly by taking $h(y)$ to be a Fourier cosine series truncated after $n+1$ terms:
\begin{align}
   h(y) & = \frac{a_0}{2} + \sum_{k=1}^n a_k \cos(k\, y) \label{eq-fourierseries}.
\end{align}
One could envisage using other representations but, with the possible exception of
the differential sedimentation kernel, this simple approach will suffice for our purposes. Once the exponents $\mu$ and $\nu$ have been determined, the inverse problem reduces to using the observed size distributions to determine the coefficients of these Fourier series. The relative merits of these two approaches will be discussed below.

\section{The forward problem}
In this section we describe a few features of the solution of the forward problem which are relevant to our subsequent discussion of the inverse problem in the sense that we use numerical solutions of Eq.\eqref{eq-Smoluchowski} as input. We mention both the time-dependent and stationary cases.

\subsection{Time-dependent case}
\label{sec:timeDependentForwardProblem}
\begin{figure}[tb]
  \includegraphics[width=0.9\columnwidth]{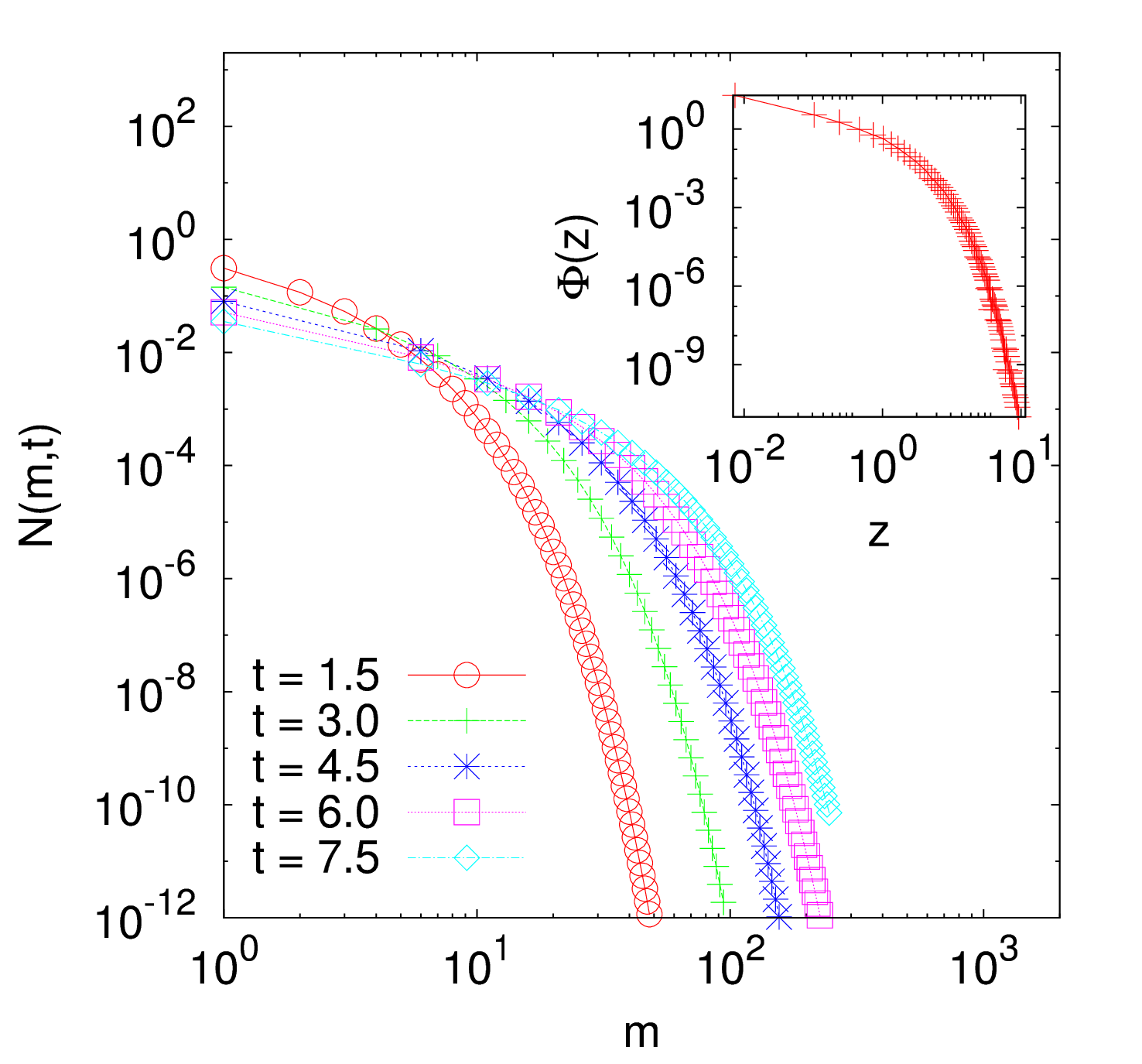}
  \caption{An example of scaling for a time-dependent solution of Eq.\eqref{eq-Smoluchowski}. The main panel shows the time evolution of the particle size distribution for the kernel $K(m_1,m_2) = \sqrt{m_1} + \sqrt{m_2}$. The inset panel shows the scaling function, $\Phi(z)$, obtained by rescaling the data according to Eq.\eqref{eq-scalingRelation}.}
  \label{fig:scalingExample}
\end{figure}

We first consider the evolution of the particle size distribution in the absence of a source of monomers. We must start from a particular initial distribution, which we usually take to be monodisperse: $N_m(0) = \delta(m-m_0)$. If $\lambda > 1$ then the typical particle size diverges in a finite time, $t_c$. This divergence leads to an apparent loss of mass from the system as material is absorbed into an infinite cluster.  This phenomenon is known as gelation \cite{DongenErnst1986}. Furthermore if $\nu>1$, then gelation occurs instantaneously in the absence of the cut-off, $M$ \cite{van_dongen_possible_1987}. In such cases, to make sense of Eq.\eqref{eq-Smoluchowski} requires careful consideration of the regularising role of the cut-off as discussed in \cite{ball_instantaneous_2011}. In order to avoid the complications of gelling systems, we shall restrict ourselves here to kernels for which $\lambda < 1$ and $\nu < 1$ so that no gelation occurs.

For homogeneous collision kernels, the time evolution of the cluster size distribution tends to become self-similar. That is to say, $N_m(t)$ tends to
the scaling form:
\begin{align}
  N(m,t) & \sim s(t)^{-2}\Phi(z)\hspace{1.0cm}z=\frac{m}{s(t)} \label{eq-scalingRelation},
\end{align}
where
\begin{align}
s(t) & = \frac{\mathcal{M}_2(t)}{\mathcal{M}_1(t)} \label{eq-SrateOfChange}
\end{align}
is the typical particle size and "$\sim$" denotes the scaling limit, $m \to \infty$ and $s(t) \to \infty$ with $z$ finite. For a modern review of the scaling theory of the Smoluchowski equation see \cite{Leyv2003}. The scaling function, $\Phi(z)$ satisfies the following equation:
\begin{align}
 \nonumber  0 &= \frac{1}{2}\int_0^{z}dz_1\, \kappa(z_1,z-z_1) \Phi(z_1) \Phi(z-z_1)\\
\label{eq-scaledSmol}  & - \Phi(z) \int_0^{\infty}dz_1\, \kappa(z,z_1)\, \Phi(z_1) + \left(2\Phi(z) + z\frac{d\Phi}{dz}\right).
\end{align}
Here, $\kappa(z_1,z_2) = K(z_1,z_2)/W $ , where $W$ is the separation constant generated by the self-similarity  ansatz \cite{DongenErnst1988,Leyv2003}. An illustrative example of the kind of data which we obtain from a numerical integration of the time-dependent forward problem is shown in Fig.~\ref{fig:scalingExample}.

\subsection{Stationary case}
\label{sec:statdists}
A stationary cluster size distribution is obtained in the limit of large times when a source and sink of particles is present. One might imagine that such a stationary state could provide a conceptual model of droplet dynamics in a cloud where small droplets formed by an ongoing condensation process are driven by air movements to collide and coalesce to form larger droplets that eventually become heavy enough to overcome updrafts and fall out of the cloud as rain. The stationary SCE in the presence of a source of monomers is:
\begin{align}
\nonumber  0 &= \frac{1}{2}\int_0^{m}dm_1\, K(m_1,m-m_1)\, N_{m_1}\, N_{m-m_1}\\
\label{eq-stationarySmol}  &- N_m  \int_0^{M}dm_1\, K(m,m_1)\, N_{m_1} + \frac{J}{m_0}\delta(m-m_0)
\end{align}
We distinguish two types of stationary solutions depending on whether the kernel has $|\nu - \mu| \leq 1$ or $|\nu - \mu| > 1 $. We refer to the former as "local'' kernels and the latter as "nonlocal'' for the reasons outlined in \cite{CRZ2004stat}. For kernels having $|\nu - \mu| < 1 $ (local), the stationary solution of Eq.\eqref{eq-stationarySmol} as $M \rightarrow \infty$ takes the power law form
\begin{align}
  N_m & = A \sqrt{J} m^{-\frac{\lambda + 3}{2}},
\end{align}
where $A$ is a constant which can be calculated explicitly \cite{Hayakawa1987,CRZ2004stat}. For kernels having $|\nu - \mu| > 1$ (nonlocal) the  solution is of the approximate form \cite{ball_collective_2012}:
\begin{align}
  N_m & \approx B \sqrt{J} M^{\left(m^{-\gamma}-1 \right)}m^{\nu}
\end{align}
where $B$ is a nonuniversal constant in the sense that it depends on $M$ and $\gamma = \nu - \mu -1$. Whether the kernel undergoes gelation or not is irrelevant to the stationary state. Therefore when we consider the stationary inverse problem, we do not need to restrict our choice of kernel to the same extent as we do for the time-dependent case. In particular we can consider the differential sedimentation kernel.

In order to generate stationary solutions of the SCE we can integrate Eq.\eqref{eq-Smoluchowski} forward in time until stationarity is achieved. This can be quite slow.  Indeed we have shown in \cite{ball_collective_2012} that for nonlocal kernels the stationary state can become unstable for large $M$. This curious result means that for certain kernels, the stationary state cannot be obtained by time integration.  For these reasons it is useful to be able to compute the stationary state directly without needing to compute the time evolution numerically. The following algorithm achieves this for the model kernel Eq.\eqref{eq-symmkernel}. The presentation follows the method outlined in the supplementary material of \cite{ball_collective_2012} and is similar to the work in \cite{DoraoJakobsen2006}. It is included here for completeness. To compress the notation it is helpful to introduce the moments, $\mathcal{M}_p $, of the size distribution:
\begin{align} 
\mathcal{M}_p &= \sum_{m=1}^{M} m^p N_m.
\end{align}
Using the discrete form of equation (\ref{eq-stationarySmol}), and a kernel of the form in (\ref{eq-symmkernel}), we can then use the moments to decompose (\ref{eq-stationarySmol}) as:
\begin{align} 
 N_m &= \frac{ G +  \mathcal{J} }{ \left( m^{\mu} \mathcal{M}_{\nu}  + m^{\nu} \mathcal{M}_{\mu} \right) },
\label{eq-momentstatSmol}   
\end{align}
where
\begin{align} 
  G &=  \sum_{m_1=1}^{m-1} K_0(m_1,m-m_1)\, N_{m_1}\, N_{m-m_1} \\
\mathcal{J} &= \frac{2J}{m_0}\delta(m-m_0).
\end{align}
Setting $m_0 = 1$ gives the stationary monomer density
\begin{align}
 N_1 &= \frac{2J}{\mathcal{M}_{\nu}  +  \mathcal{M}_{\mu}}. 
\end{align}
Given the behaviour of the equation for $G$ this permits a recursive definition of a stationary distribution, if the pair of moments $(\mathcal{M}_{\mu}, \mathcal{M}_{\nu}) $ are known.

\paragraph*{}
If $J,\,\mu$ and $\nu$ are known then (\ref{eq-momentstatSmol}) can be used to infer the rest of the stationary distribution by treating the problem as one of parameter estimation. In this case we seek the correct values of the pair of moments $(\mathcal{M}_{\mu}, \mathcal{M}_{\nu}) $ which will then generate the correct stationary state distribution. By treating $N_m$ as a function of the pair of moments $ N_m(\mathcal{M}_{\mu}, \mathcal{M}_{\nu})$ we can create an objective function $\Psi(\mathcal{M}_{\mu}, \mathcal{M}_{\nu}) $ to be minimised.
\begin{align}
\nonumber \Psi(\mathcal{M}_{\mu}, \mathcal{M}_{\nu}) &= (\mathcal{M}_{\mu} - \sum_{m=1}^M m^{\mu} N_m(\mathcal{M}_{\mu}, \mathcal{M}_{\nu}))^2  \\
 & \quad + (\mathcal{M}_{\nu} - \sum_{m=1}^M m^{\nu} N_m(\mathcal{M}_{\mu}, \mathcal{M}_{\nu}))^2 \\
(\mathcal{M}_{\mu}*, \mathcal{M}_{\nu}*) &= \arg \min_{(\mathcal{M}_{\mu}, \mathcal{M}_{\nu})} \Psi(\mathcal{M}_{\mu}, \mathcal{M}_{\nu})
\label{eq-statgenminimise}
\end{align}
We remark that this method relies on the special structure of the kernel Eq.\eqref{eq-symmkernel}. In general, one cannot avoid time integration of the SCE.

\section{The stationary inverse problem}
\label{sec:retrievestatdistkern}

We now present results for the stationary inverse problem.  The stationary particle size distributions for different kernels were obtained by numerically solving the forward problem either by time integration of Eq.\eqref{eq-Smoluchowski} or using the algorithm outlined in Sec. \ref{sec:statdists}. The objective is to use the stationary $N_m$ to reconstruct $K(m_1,m_2)$. We take the rate of mass input to be $J=1$. In principle, the value of $J$ should be obtained from the stationary size distribution. For simplicity, we assume that the value of $J$ is known a-priori. 

\subsection{Description of the method}

We split the inverse problem into two stages:
\begin{enumerate}
\item 
Fit the data to the model kernel 
\begin{displaymath}
K_0(m_1,m_2)=\frac{g}{2}\left(m_1^{\mu}m_2^{\nu} + m_2^{\mu}m_1^{\nu}  \right)
\end{displaymath}
to obtain approximate values for $g$, $\mu$ and $\nu$.
\item 
Estimate the shape function keeping $g$, $\mu$ and $\nu$ fixed at the values obtained in step 1.
\end{enumerate}

To implement the first stage, we use the observed values of $N_m$ to define
\begin{align}
\nonumber \mathcal{R}_1(m,g,\mu,\nu) &= \frac{1}{2} \sum_{m_1=1}^{m-1} K_0(m_1,m-m_1)\,N_{m_1}\,N_{m-m_1} \\
& - N_m\,\sum_{m_1=1}^M K_0(m,m_1) N_{m_1} + \delta_{m,1}.
\end{align}
From these we construct the objective function:
\begin{align}
 S_1(g,\mu,\nu) &= \frac{1}{M} \sum_{m=1}^{M} \mathcal{R}_1(m,g,\mu,\nu)^2
\end{align}
We now estimate the values of $g$, $\mu$ and $\nu$ by minimising this function:
\begin{align}
\label{eq-minimisation}   (g*,\mu^*,\nu^*) &= \arg\min_{(g,\mu,\nu)}\:S_1(g,\mu,\nu).
\end{align}
This is a nonlinear least squares problem since $\mu$ and $\nu$ enter into the objective function as exponents. We solve it numerically using the Nelder-Mead algorithm.  We used the implementation provided in \textit{Mathematica}\texttrademark in its \texttt{NMinimize} function. 

At the second stage, we fix $\mu$ and $\nu$ to the values obtained in stage one and introduce a corrected kernel:
\begin{equation}
\label{eq-combK}
K(m_1,m_2)=\frac{g^*}{2}\left(m_1^{\mu^*}m_2^{\nu^*} + m_2^{\mu^*}m_1^{\nu^*}  \right)\, f\left(\frac{m_1}{m_2}\right)
\end{equation}
where the shape function, $f(x)$, is given by Eq.\eqref{eq-hy}. It depends on $n+1$ Fourier coefficients, $\{a_k\}_{k=0}^n$, through Eq.\eqref{eq-fourierseries}.
Choosing an appropriate value for $n$ is important to get good results. If $n$ is too low, Eq.\eqref{eq-fourierseries} has insufficient flexibility to adequately represent the shape function. If $n$ is too high, we start finding implausibly oscillatory functions. The choice of $n$ represents our prior expectation of how rough or wiggling a function we expect the collision kernel to be. In all results presented below, the value of $n$ was chosen empirically to be equal to 3, giving 4 Fourier coefficients in all. We now proceed as before and define
\begin{align}
\nonumber \mathcal{R}_2(m,a_0\ldots a_n) &= \frac{1}{2} \sum_{m_1=1}^{m-1} K(m_1,m-m_1)\,N_{m_1}\,N_{m-m_1} \\
& - N_m\,\sum_{m_1=1}^M K(m,m_1) N_{m_1} + \delta_{m,1}.
\end{align}
From these we construct the objective function:
\begin{align}
 S_2(a_0\ldots a_n) &= \frac{1}{M} \sum_{m=1}^{M} \mathcal{R}_2(m,a_0\ldots a_n)^2.
\end{align}
The values of the coefficients, $a_0\ldots a_n$ are then obtained by solving the {\em linear} least squares problem
\begin{align}
\label{eq-minimisation2}   (a_0^*\ldots a_n^*) &= \arg\min_{a_0\ldots a_n}\:S_2(a_0\ldots a_n).
\end{align}
As suggested in \cite{OMTKK2011}, one could weight the objective function by the values of $N_m$, on the basis of the heuristic that regions with higher density contain more information. We did not find any considerable improvement from using such weighting so, in the interest of simplicity, all results presented in this paper are unweighted. 

\subsection{Results}

\begin{figure}[tb]
  \includegraphics[width=\figwidth]{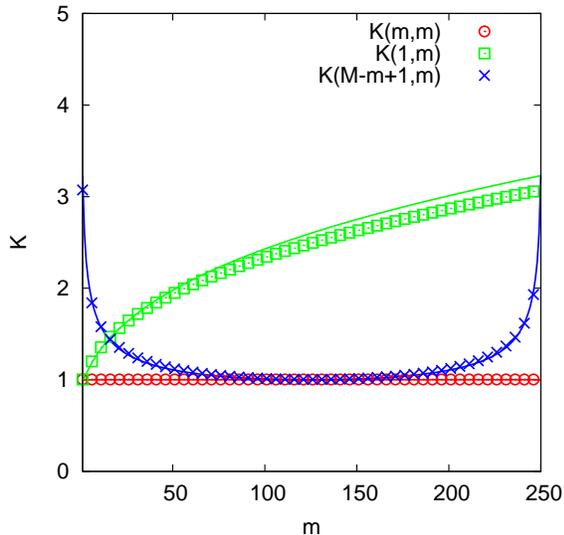}
  \caption{Numerical solution of the stationary inverse problem for the model kernel, Eq.\eqref{eq-symmkernel}, having $\nu=-\mu=1/3$. The maximum mass was $M=250$ and $n+1=4$ Fourier coefficients were used. The numerical values of the parameters were $\nu=0.3333$, $\mu=-0.3332$, $g=0.9933$, $a_0=0.9933$, $a_1=0.0268$, $a_2=-0.0153$ and $a_3=-0.0006$.}
  \label{fig:VD}
\end{figure}

\begin{figure}
\centering
\subfigure[\ Brownian coagulation kernel]{
\label{fig-BCSS}
\includegraphics[width=\figwidth]{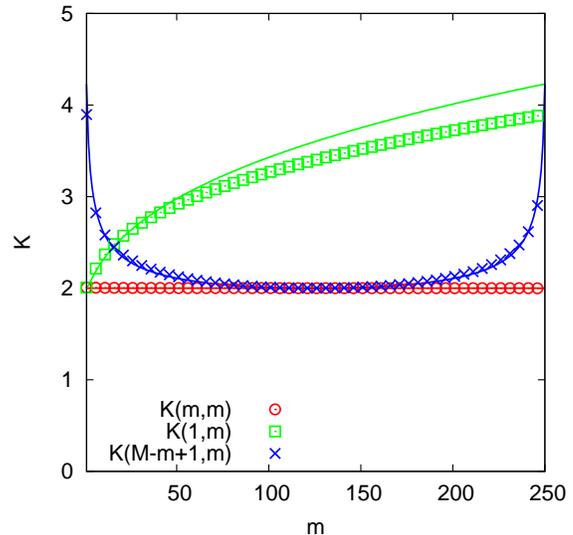}
}
\subfigure[\ Saturn's rings kernel]{
\label{fig-SRSS}
\includegraphics[width=\figwidth]{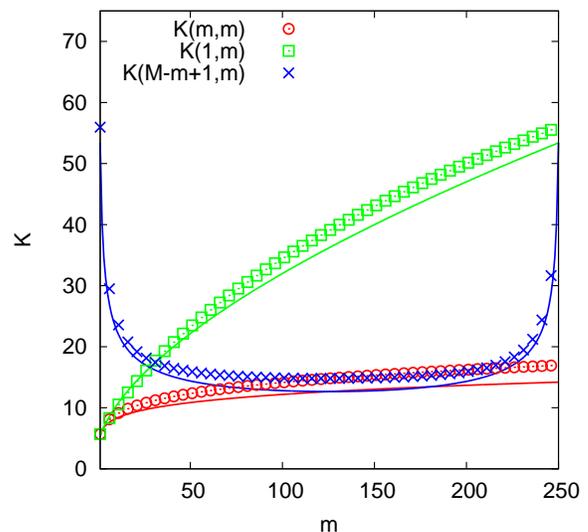}
}
\subfigure[\ Nonlinear shear kernel]{
\label{fig-NLSVSS}
\includegraphics[width=\figwidth]{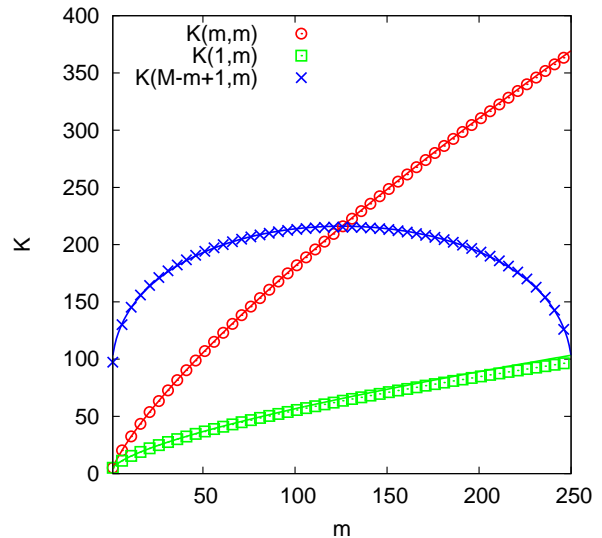}
}
\caption{Numerical solutions for the stationary inverse problems for the model kernels given in Eqs. \eqref{eq-KBC}, \eqref{eq-KSR} and \eqref{eq-KNLSV}. In all cases, the maximum mass was $M=250$ and $n+1=4$ Fourier coefficients were used}
\end{figure}

We now show some results obtained by applying this method to some stationary size distributions. In order to compare the true kernel to the kernels constructed from our inverse method we choose to plot slices through the kernels as a function of $m$ rather than to show two-dimensional surface plots. This is simply to aid clarity. We compare a slice through the diagonal, 
$K(m,m)$, a slice through the edge, $K(m,1)$ and a transverse slice, $K(M-m,m)$. Taken together these one-dimensional slices give a good overall sense of the quality of the fit. As a sanity check, we verified that this algorithm recovers a reasonable approximation to the kernel if the stationary $N_m$ is generated using a kernel of the form Eq.\eqref{eq-symmkernel} for which the shape function is unity.  Figure \ref{fig:VD} shows the results of this test for the case of Eq.\eqref{eq-symmkernel} with $\nu = -\mu = 1/3$ and a maximum mass of $M=250$. The various slices through the kernel mentioned are shown. The points representing the reconstructed values closely follow and the solid lines representing the true curves. 

We now consider the test kernels, Eq.\eqref{eq-KBC}-\eqref{eq-KNLSV} which
have nontrivial shape functions. The values of the exponents $\mu$ and $\nu$ at the first stage are usually not particularly good. For example, in the case of the Brownian coagulation kernel ($\nu=1/3$ and $\mu=-1/3$), we obtain $\nu^*\approx 0.24$ and $\mu^*\approx -0.24$. The reason is clear from Fig.\ref{fig-shapefunctions}: with a cut-off of order $10^2$, most of the data are in the central region where the shape function is strongly varying. Indeed Fig.\ref{fig-shapefunctions} suggests that one would need data with a cut-off of order $10^6$ in order to comfortably enter the asymptotic regime for of these kernels where the "true" values of $\mu$ and $\nu$ should become apparent.

We then estimated the shape function taking $n=3$ and using Eq.\eqref{eq-fourierseries} to represent the the shape function. The results were not particularly close to the true shape function although they seemed qualitatively similar. At this point it is important to recognise that neither the estimated  exponents $\mu^*$ and $\nu^*$ nor the estimated shape function are individually of any importance. What matters is the degree to which the combination of the two, as written in Eq.\eqref{eq-combK}, approximates the true kernel over the range of scales for which we have data on the stationary size distribution. It is possible for the estimated  exponents $\mu^*$ and $\nu^*$ and the estimated shape function to be individually poor approximations to the "truth" and yet provide a good approximation to the true kernel when combined. Furthermore, it is ok if the reconstructed kernel approximates the true kernel very poorly when extrapolated beyond the range of scales, $\left[1, M\right]$, since we have 
no data outside of this range.

The results for these kernels are shown in Figs. \ref{fig-BCSS}, \ref{fig-SRSS} and \ref{fig-NLSVSS} for the Brownian coagulation,  Saturn's rings  and nonlinear shear kernels respectively (see Eqs. \eqref{eq-KBC}, \eqref{eq-KSR} and \eqref{eq-KNLSV}). It is clear that the method does an excellent job of recovering an approximation to the truth in all cases. The method is therefore quite robust.

\subsection{The differential sedimentation kernel}

\begin{figure}[tb]
\includegraphics[width=\figwidth,keepaspectratio=true]{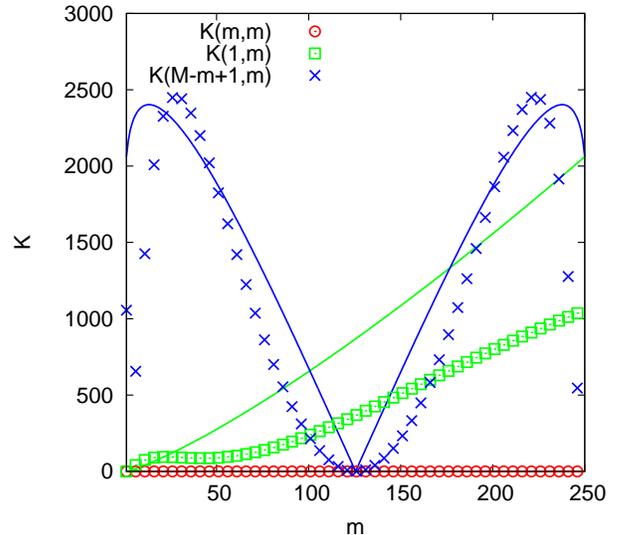}
\caption{\label{fig-DSSS} Numerical solution for the stationary inverse problem for the differential sedimentation kernel, Eq. \eqref{eq-KDS}.The maximum mass was $M=250$ and a fully nonlinear estimation algorithm with $n+1=4$ Fourier coefficients was used as described in the main text. }
\end{figure}

While all seems well at this point, when we applied the method to the differential sedimentation kernel, Eq.\eqref{eq-KDS},  most relevant to the cloud physics problem, we found that it entirely failed to reconstruct anything reasonable. In particular, we obtained negative values for the shape function if more than 2 Fourier coefficients were used. We can trace the problem to the presence of the cusp in the shape function at zero (see Fig. \ref{fig-shapefunctions}). To adequately capture this cusp using a Fourier cosine series would require the retention of a very large number of terms which, as we have already seen, is not a good idea since it provides too much freedom to introduce extraneous oscillations. 

In order to get some reasonable results for this kernel, after some experimentation, we combined steps 1 and 2 of the previous method into one single optimization problem. This is considerably more expensive numerically since one now has to do a fully nonlinear minimization of the objective function in the six-dimensional space ($\nu, \mu, a_0, a_1, a_3, a_3)$. In addition it was necessary to include explicit constraints to prevent the shape function from becoming negative near 1 which further
slow down the calculation. The results are shown in Fig. \ref{fig-DSSS}. While the algorithm recovers the correct qualitative form of the kernel, we see there are large quantitative differences compared to the results for the other kernels shown in Figs.\ref{fig-BCSS}, \ref{fig-SRSS} and \ref{fig-NLSVSS}. Further research will be required to improve the performance and reliability of the method for such cases.

Had it not been for the fact that we already knew the form of the solution, we probably would not have been able to reconstruct the kernel even in the approximate way shown in Fig. \ref{fig-DSSS}.  This reinforces our belief, stated at the outset, that the kind of data-driven approach demonstrated above can be complementary to existing theoretical and numerical studies of coagulation phenomena but clearly cannot replace them. To conclude our discussion of the differential sedimentation kernel, we remark that the problems we have encountered due to the cusp in the shape function are unlikely to occur if gravitational sedimentation is occuring in a turbulent environment like a cloud. This is because spatial variations in the turbulent velocity field provide an additional mechanism for collisions between droplets of equal size
as originally pointed out by Saffman and Turner \cite{saffman_collision_1956}. This leads to a smoothing of the cusp and a nonzero value for the collision rates of equally sized droplets. In \cite{OMTKK2011} it was shown that a linear combination of
the differential sedimentation and Saffman-Turner kernels is a plausible model in
the case of gravitational settling in turbulent environments.

\section{The time-dependent inverse problem}
\label{sec:scalingDecayRetrieval}

\begin{figure}[tb]
\includegraphics[width=\figwidth]{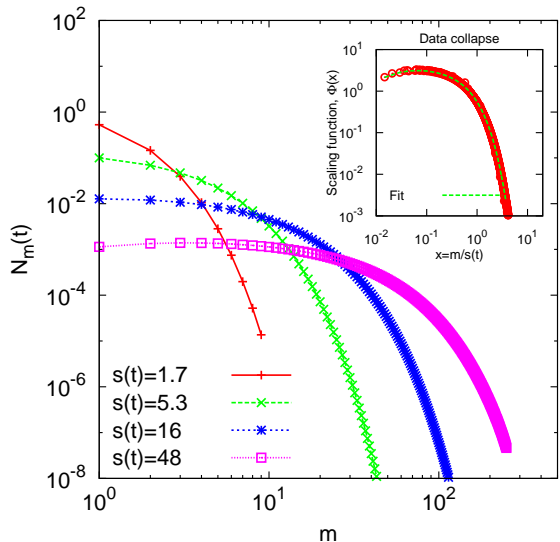}
\caption{Input data for the time-dependent inverse problem in the case of the Brownian coagulation kernel, Eq.\ref{eq-KBC}. Main panel shows the time evolution of the size distribution and the inset shows the data collapse obtained by rescaling this data according to Eq.\eqref{eq-scalingRelation}.}
\label{fig-decayInput} 
\end{figure}

In the case of the time-dependent inverse problem, some modifications of the method described above permits the reconstruction of the collision kernel from a succession of snapshots of the particle size distribution provided that the data span a sufficient range of mass and time scales to enter into the scaling regime described in section \ref{sec:timeDependentForwardProblem}. The basic idea is to use the scaling ansatz, Eq.\eqref{eq-scalingRelation}, to collapse
the different snapshots of the particle size distribution onto a single scaling function, $\Phi(z)$. This curve satisfies Eq.\eqref{eq-scaledSmol}. Given that $\Phi(z)$ is known, the corresponding inverse problem for $\kappa(z_1,z_2)$  is structurally almost identical to the stationary inverse problem which we have already discussed in Sec.\ref{sec:retrievestatdistkern}. After appropriate discretisation, of Eq.\eqref{eq-scaledSmol}, the methods described in Sec. \ref{sec:retrievestatdistkern} can be applied with some minor modifications of the objective functions $S_1(\mu,\nu)$ and $S_2(a_0\ldots a_n)$ to take into account the additional linear terms in Eq.\eqref{eq-scaledSmol}.

The steps in the procedure are as follows
\begin{enumerate}
\item
From the observed snapshots of the size distribution, $N_m(t)$ we calculate the characteristic particle size as given by Eq.\eqref{eq-SrateOfChange} and use these values to rescale the data according to Eq.\eqref{eq-scalingRelation}. Provided that our measured size distributions are in the scaling regime, this should collapse the data onto a single scaling curve, $\Phi(z)$. This is done for the Brownian coagulation kernel, Eq.\eqref{eq-KBC} in Fig. \ref{fig-decayInput} with the data collapse shown in the inset.
\item
To discretise Eq.\eqref{eq-scaledSmol}, we need to calculate $\Phi(z)$ and its derivative on a regular grid. It is therefore convenient to fit the collapsed data to a specific functional form. The fit was done using regression in the logarithmic variables $(u=\log.  \Phi,  v=\log z)$ by fitting the
the data to the model
\begin{displaymath}
u = b_0 + b_1 v + b_2 v^2 + b_3v^3 + b_4 \exp(v)
\end{displaymath}
and then recovering the required curve by exponentiation. The result of this fit is superimposed on the data in the inset of Fig.\ref{fig-decayInput}. The fitted curve can then be differentiated analytically to compute the lefthand side of Eq.\eqref{eq-scaledSmol}.
\item
We discretise Eq.\eqref{eq-scaledSmol} on $Z$ uniformly spaced points, $\{z_i = z_{\text{min}} + (i-1)\,\Delta z\}_{i=1}^Z$ where 
\begin{displaymath}
z_i = z_{\text{min}} + (i-1)\,\Delta z
\end{displaymath}
and
\begin{displaymath}
\Delta z = (z_{\text{max}}- z_{\text{min}})/(Z-1).
\end{displaymath}
$z_{\text{min}}$ is set by the maximum size reached by $s(t)$. We choose $z_{\text{max}}$ so that it lies in the exponenial tail of the scaling function (see inset of Fig.\ref{fig:scalingExample}) but not so large that the $\Phi(z)$ is effectively negligible. In the results presented below we typically took $Z=250$ as the number of discretisation points.
\item 
We now proceed as before by defining
\begin{align}
\nonumber \mathcal{R}_1(i,g,\mu,\nu) &= \frac{1}{2} \sum_{j=1}^{i-1} K_0(z_j,z_i-z_j)\,\Phi(z_j)\,\Phi(z_i-z_j)\,(\Delta z) \\
& - \Phi(z_i)\,\sum_{j=1}^Z K_0(z_i,z_j) \Phi(z_j)\, (\Delta z)\\
\nonumber &+ 2\Phi(z_i) + z_i\frac{d\Phi}{dz}(z_i)
\end{align}
We again construct the objective function:
\begin{align}
 S_1(g,\mu,\nu) &= \frac{1}{Z} \sum_{i=1}^{Z} \mathcal{R}_1(i,g,\mu,\nu)^2 
\end{align}
and obtain the estimated values of $g$, $\mu$ and $\nu$:
\begin{align}
\label{eq-minimisationDecay}   (g^*,\mu^*,\nu^*) &= \arg\min_{(g,\mu,\nu)}\:S_1(g,\mu,\nu).
\end{align}
\item
Finally we correct this result with a shape function:
\begin{equation}
\label{eq-combKappa}
\kappa(z_1,z_2)=\frac{g^*}{2}\left(z_1^{\mu^*}z_2^{\nu^*} + z_2^{\mu^*}z_1^{\nu^*}  \right)\, f\left(\frac{z_1}{z_2}\right)
\end{equation}
where the shape function, $f(x)$, is again given by Eq.\eqref{eq-hy} with the $n+1$ Fourier coefficients, $\{a_k\}_{k=0}^n$, entering  via Eq.\eqref{eq-fourierseries}. We now proceed as before and define
\begin{align}
\nonumber \mathcal{R}_2(i,a_0\ldots a_n) &= .\frac{1}{2} \sum_{j=1}^{i-1} \kappa(z_j,z_i-z_j)\,\Phi(z_j)\,\Phi(z_i-z_j)\,(\Delta z) \\
& - \Phi(z_i)\,\sum_{j=1}^Z \kappa(z_i,z_j) \Phi(z_j)\, (\Delta z)\\
\nonumber &+ 2\Phi(z_i) + z_i\frac{d\Phi}{dz}(z_i)
\end{align}
From these we construct the objective function:
\begin{align}
 S_2(a_0\ldots a_n) &= \frac{1}{Z} \sum_{i=1}^{Z} \mathcal{R}_2(i,a_0\ldots a_n)^2 
\end{align}
and solving the least squares problem to obtain the coefficients:
\begin{align}
\label{eq-minimisationDecay2}   (a_0^*\ldots a_n^*) &= \arg\min_{a_0\ldots a_n}\:S_2(a_0\ldots a_n).
\end{align}
\end{enumerate}
This procedure  gives the rescaled kernel, $\kappa(z_1, z_2)$. To return to the original scale requires a knowledge of the separation constant, $W$, which enters when using the self-similarity ansatz to split the time-dependent problem into an equation for $s(t)$ and a time-independent equation for $\Phi(z)$. The presence of this arbitrary constant reflects the fact that a rescaling of the amplitude of the solution corresponds to a rescaling of time. If we wish to fix a value of this constant, we need a way to set the time scale. This can be done by fitting the data curve obtained using (\ref{eq-SrateOfChange}) with the analytic solution for $s(t)$:

\begin{align}
   s(t) &= \left[(1-\lambda)Wt + X  \right]^{\frac{1}{1-\lambda}}, \quad \lambda < 1
\end{align}

\paragraph*{}
Here $X$ is a parameter replacing the initial condition term $s(0)^{1-\lambda}$. Provided that $(\mu, \nu)$ are retrieved sufficiently well during the estimation of the kernel, then the homogeneity $\lambda=\mu+\nu$ is known and the fitting process to obtain $W$ works well. Multiplying the retrieved kernel estimate $\kappa(z_1,z_2)$ by $W$ rescales the result to match the original unscaled input kernel $K(z_1, z_2)$.

\begin{figure}[tb]
\includegraphics[width=\figwidth]{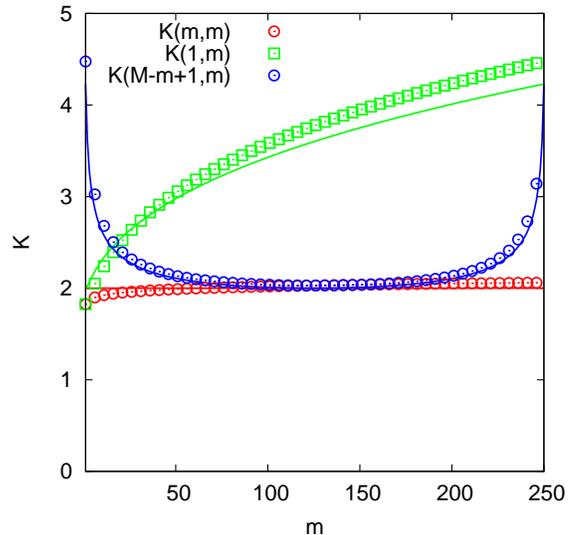}
\caption{Solution of the time-dependent problem with Brownian coagulation kernel, Eq.\eqref{eq-KBC}, based on the data collapse shown in Fig.\ref{fig-decayInput}. The number of discretisation points used was $Z=250$ and $n+1=4$ Fourier coefficients were used. The numerical results have been rescaled by a factor of 1.8 coming from the separation constant (see main text).}
\label{fig-decayResults} 
\end{figure}

Some illustrative results from the application of this method to some time-dependent inverse problems for the case of the Brownian coagulation kernel, Eq.\eqref{eq-KBC},  are shown in Fig.\ref{fig-decayResults}. 
These calculations were done with $n=3$. Although we get excellent results in this case, in general the results tend not to be as good as in the stationary case. One reason for this is the increased error made by assuming that the system is in the scaling regime. In no case was the data collapse perfect.  A second reason is that the exponenially decaying tails of the scaling function likely put fewer constraints on the form of the kernel since the size distribution becomes negligibly small there.

\section{Conclusions and Outlook}
\label{sec:conclusionsAndOutlook}
To conclude, we have presented an approach to solving the inverse problem of reconstructing the collision kernel from observations of the particle size distribution for an ensemble of irreversibly coalescing particles whose statistical dynamics is modeled by the Smoluchowski coagulation equation. Compared with previous work on this problem, our approach provides a lot of flexibility in the functional form of the collision kernel. In particular, it handles the possibility of fractional powers of particle masses in the kernel, a situation which occurs commonly in applications, in an elegant way. We applied our method to a selection of stationary and time-dependent inverse problems for which the size distributions were obtained by numerically solving the forward Smoluchowski problem with a variety of different collision kernels taken from various branches of physics. The results were of sufficiently high quality to demonstrate the feasibility of using this kind of data-driven approach to reconstruct collision 
kernels from data.

Since this problem is under-determined, some prior expectation of the kind of functional forms which are reasonable for the collision kernel is required in order to obtain high quality results from our method. This prior expectation is encoded in the choice of the number of Fourier coefficients, $n$, with which to represent the shape function of the collision kernel. The choice of $n$ reflects the degree of wiggling which we think is plausible. It is probably possible to come up with cross-validation arguments to help to select the value of $n$ but it is not possible to remove this arbitrariness entirely. For this reason, our approach complements rather than replaces existing direct theoretical and numerical approaches to coagulation phenomena. It is likely to be most useful in situations  for which quality measurements of the particle size distribution are available but for which the underlying microphysics is unknown or controversial. 

In terms of future research, it would be interesting to apply these techniques to some real data. It will therefore be necessary to quantify how well things work in the presence of observational noise. It will also be necessary to quantify the uncertainty in the collision kernels which are reconstructed from the data. To do this, it may be best to reformulate the problem in a probabilistic setting and apply some of the methods of Bayesian inverse problems which have been developed recently (see for example \cite{cotter_bayesian_2009}). Such a reformulation would also allow us to be more explicit about the prior information about the kernel which is incorporated into the model and provide a means for observational data to over-rule these prior choices if they are inconsistent with the observations. 

Finally, the central challenge in the inverse problem is that the given information is a function of one variable, whereas the kernel is a priori a function of two variables.  We have shown that constraining K to be homogeneous  renders the inversion tractable in representative simple cases. In doing so we find ourselves using only the steady state and/or scaling forms of the cluster size distribution.  In principle one could exploit the full time dependence of the cluster size distribution.   This could enable K to be inferred as a function of two masses without assuming homogeneity. We hope to return to this in future work.

\begin{acknowledgments}
C.C. thanks R. Onishi for discussions and for providing some important references and acknowledges the
support of the EPSRC (grant EP/H051295/1) and the EU COST Action MP0806, Particles in Turbulence.
\end{acknowledgments}


%

\end{document}